\documentstyle[prl,aps,preprint,tighten,floats]{revtex}

\newcommand{\ket}[1]{\left|#1\right\rangle}

\begin{document}
\draft

\preprint{\vbox{Foundations of Quantum Physics\null\hfill\rm\today}}

\title{A Realizable, Non-Null\\Schr\"{o}dinger's Cat Experiment}

\author{Martin~S.~Altschul\footnote{altschulma@kpnwoa.mts.kpnw.org}}
\address{Northwest Permanente \\
Salem, Oregon 97306}

\author{Brett~D.~Altschul\footnote{baltschu@mit.edu}}
\address{Department of Physics \\
Massachusetts Institute of Technology \\
Cambridge, Massachusetts 02139}

\maketitle

\thispagestyle{empty}

\begin{abstract}%
Working from the Schr\"{o}dinger's Cat paradigm, a series of experiments are
constructed. The Bedford-Wang experiment is examined, and the ambiguity in its
meaning is addressed. We eliminate this ambiguity by abandoning the idea of the
triggering event, replacing the two-state system with a mirror that undergoes
wave packet spreading. This creates an experimentally testable version of a
modified Schr\"{o}dinger's Cat experiment for which a null result is not the
obvious outcome.

\end{abstract}

\narrowtext
\newpage

\section{Introduction}
\noindent
Twenty years ago Bedford and Wang (BW) \cite{ref-bw1,ref-bw2} devised an
experimentally realizable
version of the Schr\"{o}dinger's Cat (SC) experiment. They started with an
ordinary double slit experiment using photons of wavelength $\lambda_{1}$; they
then added a refinement. The slits have movable
slit covers that allow two possible configurations:
\begin{quote}
1. Configuration Ab has slit A open and slit B closed.\\
2. Configuration aB has slit A closed and slit B open.
\end{quote}
\noindent
The slit control system is triggered by photocells registering the
output of a beam splitter. In this way, the slit system (SS) can be set up so
that its state vector is entirely determined by a single photon processed by
the beam splitter/photocell system. BW claim that if the beam splitter outputs
a photon in a 50/50 superposition state, then further application of the
superposition principle (SP) according the the standard interpretation of
quantum mechanics (SIQM) forces us to conclude that the slit system is in the
state
\begin{equation}
\ket{\psi} = \frac{1}{\sqrt{2}}(\ket{Ab}+\ket{aB}),
\end{equation}
\noindent
and consequently, a double
slit interference pattern is to be expected (Figure~\ref{BWexpt}).

BW then add that ``although the result seemed a foregone conclusion'' the
experiment was performed and yielded a null result---no double slit
interference. But is this result interesting or informative? The BW detractors
\cite{ref-guthk} claim that BW have merely misquoted and misused the
SP---making their experiment pointless. BW insist that their experiment exposes
a flaw in SIQM \cite{ref-bw3,ref-bw4,ref-bw5}.

Where do BW and their detractors agree? They agree that the null result of the
experiment is a foregone conclusion. Why do they agree? Because, it is obvious
that the position of the slit covers cannot really be uncertain in any 
quantum-mechanical sense. Their positions are ``given away'' by several easily
measured phenomena, the most obvious being thermal radiation.

Where do BW and their detractors disagree? At the core of the dispute is the
question of whether the experimenter can prevent state vector reduction by
ignoring information.  That is, by choosing not to perform obvious
measurements.  There are numerous recent examples of experiments in which
interesting effects are obtained by choosing not to measure state vectors at
some intermediate point in the experiment
\cite{ref-lwang1,ref-lwang2,ref-lwang3,ref-ou1,ref-ou2}.
But it is obvious that one cannot choose to ignore
arbitrarily large numbers of photons.  It is unclear whether SIQM provides a
prescription for deciding how many and what kind of photons (or other
particles) can legitimately be ignored.

There is another line of inquiry suggested by the BW experiment. We propose to
devise an experimental configuration in which the state vector of the movable
slit covers (or their equivalent) is not given away by thermal radiation or
any other effect, and the superposition effect we are trying to measure is not
swamped by extraneous phenomena. These conditions could produce a non-null
experimental outcome.

\section{Modified BW Experiment}
\label{gedanken}

\noindent
Our objective at this stage is to use the BW setup to develop the tools and
language to analyze the conjectured non-null experiment. We add to the BW
apparatus a source of short wavelength radiation ($\lambda_{2}$) that can be
used to probe the state (A open versus A closed) of the system, as shown in
Figure~\ref{prelim}.

If we just run the $\lambda_{1}$ part of the experiment, and if we take the
superposition state $\frac{1}{\sqrt{2}}(\ket{Ab}+\ket{aB})$ at face value then
we get a $\lambda_{1}$ double-slit pattern.  Next we observe the following
sequence:
\begin{quote}
1. Turn on the $\lambda_{1}$ source and observe the double-slit pattern. Turn
on the $\lambda_{2}$ source. (For clarity and simplicity, we can choose to use
just one $\lambda_{2}$ particle.)  As soon as D$_{\lambda_{2}}$ either detects
or fails to detect the $\lambda_{2}$ particle, the $\lambda_{1}$ double-slit
pattern must vanish.
\end{quote}
If we were to turn on the $\lambda_{1}$ source only after the $\lambda_{2}$
particle has encountered (or failed to encounter D$_{\lambda_{2}}$, then
obviously there will be no $\lambda_{1}$ double slit-pattern.

Now we introduce the following wrinkle:
\begin{quote}
1$'$. First turn on E$_{\lambda_{2}}$, but instead of placing D$_{\lambda_{2}}$
just beyond the A slit, allow the $\lambda_{2}$ particle to follow a long path
to a distant mirror and then bounce back to be detected in the lab (Figure~\ref
{pathdelay}). (Note that we can set this up for both the transmitted and
reflected paths or just one or the other.) Turn on  E$_{\lambda_{1}}$, while
the $\lambda_{2}$ particle is in flight to the distant mirror(s). Now we ask,
``Does a $\lambda_{1}$ double-slit pattern appear while the $\lambda_{2}$
particle is in flight?''
\end{quote}
Before answering, let us review the time-line of the experiment, as shown in
Figure~\ref{timeline}.
\begin{quote}
$t_{0}$---$\lambda_{2}$ emission\\
$t_{1}$---$\lambda_{2}$ encounter with SS\\
$t_{2}$---$\lambda_{1}$ emission\\
$t_{3}$---$\lambda_{1}$ encounter with SS\\
$t_{4}$---$\lambda_{1}$ encounter with double slit interference screen\\
$t_{5}$---$\lambda_{2}$ reflection\\
$t_{6}$---$\lambda_{2}$ detection
\end{quote}

Now suppose that a $\lambda_{1}$ double-slit pattern is observed at $t_{4}$.
A $\lambda_{2}$ detection event at $t_{6}$ tells us that the A slit was either
open or closed (not in the state $\frac{1}{\sqrt{2}}(\ket{Ab}+\ket{aB})$) at
time $t_{1}$, so that $\lambda_{1}$ double-slit interference cannot
occur at time $t>t_{1}$. This is not a paradox---it is a flat out
contradiction.

Now suppose that a double-slit pattern is not observed at $t_{4}$. The
experimenter (who is presumably none other than Wigner's friend) can then
choose during the interval $t_{4}<t<t_{6}$ to deactivate the $\lambda_{2}$
detectors. If the $\lambda_{2}$ particle floats away without being detected,
then the $\lambda_{1}$ double-slit pattern {\em should} be observed---again a
contradiction.

\pagebreak
For completeness, we need to dispose of one other point. The reader has
probably thought of something like the following rejoinder:
\begin{quote}
State vector reduction does not occur until time $t_{6}$; therefore, at time
$t_{2}$, the $\lambda_{1}$ quanta have equal probability of passing through
A or B and can produce an interference pattern without contradicting detection
of the $\lambda_{2}$ particle at D$_{\lambda_{2}}$ at time $t_{6}$.
\end{quote}
We can refute this rejoinder with a simple example. Suppose that particle
P$_{1}$ is subjected to a 50/50 quantum bifurcation.  It's wave function
becomes
\begin{equation}
\ket{{\rm P}_{1}}=\frac{1}{\sqrt{2}}(\ket{{\rm P}_{1}}_{+}+\ket{{\rm P}_{1}}
_{-}).
\end{equation}
Following the rejoinder, one can then argue that a second particle P$_{2}$ can
be scattered off $\ket{{\rm P}_{1}}_{+}$ even if a later measurement finds
P$_{1}$ in $\ket{{\rm P}_{1}}_{-}$.

Clearly, the rejoinder in this form is neither a sound argument, nor a proper
expression of SIQM.  Nevertheless, we can see a glimmer of something deeper if
we continue in the direction the rejoinder leads us. It seems that the
$\lambda_{1}$ and $\lambda_{2}$ observations should yield mutually compatible
sets of basis vectors for the same Hilbert space. But the usual linear
mapping works only in one direction, $\ket{\lambda_{1}}_{interference}=\frac
{1}{\sqrt{2}}(\ket{\lambda_{2}}_{A\, open}+\ket{\lambda_{2}}_{B\, open})$.
There are no expressions
\begin{eqnarray}
\ket{\lambda_{2}}_{A\, open}=\ket{\lambda_{1}}_{interference}+{\rm other
\, vectors},\nonumber\\
\ket{\lambda_{2}}_{B\, open}=\ket{\lambda_{1}}_{interference}+{\rm other
\, vectors}, \nonumber
\end{eqnarray}
a fact to which we will return in the Conclusions (Section~\ref{concl}).

Now we need to come up with a more realistic superposition state and see what
happens to this contradiction.

\section{Interference From a Mesoscopic Mirror}
\noindent
We start with a mesoscopic mirror whose wave function is
bifurcated. (That is, the mirror appears in one of two possible positions with
a 50/50 probability. The wave function for positions in between is zero.)

Both the $\lambda_{1}$ and $\lambda_{2}$ photons are reflected by the mirror.
This interval $t_{3}-t_{1}$ must be kept very short, so that the recoil of the
mirror from the $\lambda_{2}$ impact is minimized.
Figure~\ref{bifur} shows the experimental setup with the two possible positions
of the mirror separated by the distance $\frac{1}{4}\lambda_{1}$.  It is
evident from Figure~\ref{bifur} that there is an interference node for the
reflected $\lambda_{1}$ photons. We therefore can develop the same
kind of consistency argument as in Section~\ref{gedanken}.

But is it really possible to prepare the mirror in such a state? We could use a
half-silvered mirror, which bifurcates photon wave functions and so can in
principle bifurcate its own wave function by interacting with a single photon.

\pagebreak
The uncertainty in the mirror velocity from single photon bifurcation (SPB)
with photon wavelength $\lambda$ is
\begin{equation}
\Delta v_{SPB}=\frac{h}{\lambda M}.
\end{equation}
Let us compare this with the velocity uncertainty due to wave packet spreading.
In the initial state of the mirror, it is trapped with a small Gaussian
uncertainty $\Delta q_{1}$. If the mirror is then released and its wave
function allowed to spread, the characteristic initial spreading velocity
$\Delta v_{i}$ is
\begin{equation}
\Delta v_{i}=\frac{h}{4\pi\Delta q_{i}M},
\end{equation}
so
\begin{equation}
\frac{\Delta v_{SPB}}{\Delta v_{i}}=\frac{4\pi\Delta q_{i}}{\lambda}.
\end{equation}
In our opinion, in a realistic experimental setup, $\lambda\gg\Delta q_{i}$ and
hence $\Delta v_{i}>\Delta v_{SPB}$.

For our purposes, it is more difficult to work with the Gaussian distribution
associated with $\Delta v_{i}$ than to work with a bifurcated state, but it is
not impossible. Note incidentally that if we work with $\Delta v_{i}$, we have
no need of a triggering event.

Let us confine the mirror in a potential well $U$ (see the appendix for the
mechanism of the well) whose center is at $z=0$ so that for suitably small $z$,
$U$ is approximated by $U=\frac{1}{2}kz^{2}$. This gives us a harmonic
oscillator with mass $M$ (the mass of the mirror). The essence of our scheme is
to trap the mirror in the well and then dissipate energy until the ground state
of the oscillator is reached. We then have
\begin{equation}
\frac{1}{2}k(\Delta q_{i})^{2}=\frac{1}{2}M(\Delta v_{i})^2,
\end{equation}
where $\Delta q_{i}$ and $\Delta v_{i}$ are the initial position and velocity
uncertainty at the start of the experiment. We begin the experiment by turning
off $U$, so that the center of mass wave function begins to spread with
characteristic velocity $\Delta v_{i}$. We must demonstrate that
\begin{quote}
1. $\Delta v_{i}$ is really the dominant effect, and\\
2. the $\lambda_{1}$ and $\lambda_{2}$ inconsistency argument can be brought to
bear without using the especially advantageous bifurcation superposition of the
mirror wave function.
\end{quote}

\section{Experimental Geometry}
\noindent
We must now lay out the geometry that will lead to the kind of inconsistency
we are seeking. We will perform the analysis to first order; that is,
neglecting the small amplitude attenuation due to differences in the
propagation distance.

Consider a double slit experiment for $\lambda_{1}$ (Figures~\ref{pl-fig1}
and~\ref{pl-fig2}), in which the slits and the interference screen/detection
system are each $\frac{5}{2}\lambda_{1}$ from the mirror. The distance between
the slits is set at $2.29\lambda_{1}$, so that the distance between the right
and left first nodes is also $2.29\lambda_{1}$. The mirror is circular, with
diameter $\frac{5}{2}\lambda_{1}$.

We allow the wave function of the mirror to spread until $\Delta q=
\lambda_{1}$. The path length difference (PL$\Delta$) between light from the
far slit and the near slit is $\frac{1}{2}\lambda_{1}$ at the first node
when the mirror is in its initial position ($z_{m}=0$). We must compare this
with the PL$\Delta$ for the mirror positions $z=\pm\lambda_{1}$. The results
are given by Table~\ref{pl-table}.

\begin{table}
\begin{tabular}{ l|l|l|l }
$z_{m}$& $+\lambda_{1}$& $0$& $-\lambda_{1}$\\\hline
PL$\Delta$& $0.77\lambda_{1}$& $0.50\lambda_{1}$& $0.36\lambda_{1}$\\
\end{tabular}
\caption{Values of PL$\Delta$ for different mirror positions.}
\label{pl-table}
\end{table}

We see from Table~\ref{pl-table} that for $z_{m}=\pm\lambda_{1}$, the position
of the first node is substantially shifted. If the reflected $\lambda_{1}$
wave functions are based on superpositions of light reflected by the mirror
from the positions smeared out over a $\Delta q=\lambda_{1}$ Gaussian
distribution, the nodes at $x=\pm 1.145\lambda_{1}$ will be measurably less
distinct than if $z_{m}$ is fixed at $0$. {\em This is the crucial effect that
we need to observe}.

The third step in the setup of our experiment is the introduction of the
$\lambda_{2}$ photons. As in Section~\ref{gedanken}, the $\lambda_{2}$ photons
are used to measure the mirror's position. The $\lambda_{2}$ quanta, in plane
wave form, pass through an aperture of width $W_{a}$ and are incident on the
mirror at a shallow angle $\theta$. They are reflected at the same angle to the
D$_{\lambda_{2}}$ detector. Setting $W_{a}=4\lambda_{2}\cos\theta$, we see that
a detection event at D$_{\lambda_{2}}$ determines that the mirror position was
in the range
\begin{equation}
|z_{m}|\leq\frac{W_{a}}{4\cos\theta}=\lambda_{2}.
\end{equation}
If we set $\lambda_{2}=\frac{1}{4}\lambda_{1}$, then
\begin{equation}
|z_{m}|\leq\frac{1}{4}\lambda_{1}.
\label{zlimit}
\end{equation}
This condition leads to a much narrower range of values for PL$\Delta$ than
those given in Table~\ref{pl-table}. If (\ref{zlimit}) is satisfied, the
interference node is then resharpened (Figure~\ref{graphs1}). This resharpening
then leads to the same inconsistency as in Section~\ref{gedanken}.
Figure~\ref{real} shows the complete apparatus for this modified BW experiment
based on a mesoscopic mirror with wave packet spreading.

\section{Interference Patterns}
\label{actual-int}
\noindent
The interference effects on which this experiment is based are rather subtle,
so they must be treated carefully. Coherent light emerges from two pointlike
slits, A and B. The photons rebound off the wave function of the mirror and
interfere on a screen between the slits. The details of the interference
pattern vary, depending upon the mirror's wave function.

The $E$ field for the light from slit A at a point $D$ away from that slit
is given by
\begin{equation}
E_{A}=K\int P(z)\cos\left[\frac{4\pi}{\lambda_{1}}\sqrt{\left(z+\frac{5}{2}
\lambda_{1}\right)^{2}+D^{2}/4}\,\right]dz,
\end{equation}
where $P(z)=|\psi(z)|^{2}$ is the probability density for the mirror's
position and $K$ is a scaling factor. The separation between the two slits
(which should be close to 2.29$\lambda_{1}$) is $S$. For slit
B, the expression is similar,
\begin{equation}
E_{B}=K\int P(z)\cos\left[\frac{4\pi}{\lambda_{1}}\sqrt{\left(z+\frac{5}{2}
\lambda_{1}\right)^{2}+(S-D)^{2}/4}\,\right]dz.
\end{equation}

The probability density is $P(z)=\delta(z)$ if
the mirror's position is fixed. For the Gaussian wave packet,
the probability density is
\begin{equation}
\label{pgauss}
P(z)=\sqrt{\frac{1}{2\pi\lambda_{1}^{2}}}e^{-z^{2}/2\lambda_{1}^{2}}.
\end{equation}

The magnitude of the observed interference pattern is just given by the
intensity,
\begin{equation}
\label{intensity}
I=\frac{1}{2}(E_{A}+E_{B})^{2}.
\end{equation}
Since both $E_{A}$ and $E_{B}$ are functions of $D$, $I$ is a function
of $D$ as well.  The variation in $I$ in the range $0<D<S$ gives us the
interference effect.

Figure~\ref{graphs1} shows the actual interference patterns generated by two
different wave functions, with the slit separation set to $S=2.3\lambda_{1}$.
In the first, the wave function is the Gaussian (\ref{pgauss}).  In the second,
that wave function has been truncated to the region $|z_{m}|\leq\frac{1}{4}
\lambda_{1}$. The interference fringes at the edges are much sharper for the
truncated wave function, as is needed to produce the contradiction.

We have so far treated the experimental geometry as if $z$ motion were the only
wave packet spreading that occurs. There is other movement that can spoil the
experiment if not dealt with. First, let us consider sideways drift of the
mirror in the $xy$-plane. Conceptually, the simplest way to deal with this is
to have an ``out of position'' sensing system made of beams and detectors. We
then make a large number of experimental runs and use only the data obtained
when the mirror is not ``out of position.''

More difficult is tilting of the mirror out of the $xy$-plane. There are two
methods for dealing with this. The preferred method would be gyroscopic
stabilization by rotation in the $xy$-plane. This must certainly be used in the
process of trapping and confining the mirror (Appendix~A) that precedes the
experiment proper. But we need to be careful about this, as too rapid motion
could cause difficulties. The second method involves a beam and detectors. The
beam is incident perpendicular to the mirror's initial position and $\lambda
_{beam}\ll\lambda_{1}$, so, regardless of interference or reduction, the beam
provides negligible information about $z_{m}$. The recoil velocity of the
mirror from the beam impact is relatively large, but as with $\lambda_{2}$, the
beam impact is timed just before $t_{3}$, so the recoil distance is small.

\section{Restrictions on Experimental Conditions}
\noindent
We have discussed the geometry of the $\lambda_{1}$, $\lambda_{2}$
inconsistency using only the relative values of $\lambda_{1}$, $\lambda_{2}$,
and $W$. We must now demonstrate that this is the dominant effect for some
values of $\lambda_{1}$, $M$, and $\Delta q_{1}$. We start by considering the
problem of {\em radiation} from the mirror. This can ruin the experiment by
``giving away'' the mirror's position.

\pagebreak
In order to get numerical results, we have used parameters for Vanadium:
\begin{quote}
density $=\rho = 6.1$ g/cm\\
speed of sound $=v_{s}=3\times 10^{5}$ cm/s\\
atomic weight $=a_{w}=50$ amu\\
lattice spacing $=d=2.4\times 10^{-8}$ cm.
\end{quote}
More importantly, we have to choose a value for $\lambda_{1}$. The one that
seems to work best is $10^{-6}$ cm. For photons, this would mean x-rays. At
x-ray wavelengths, simple geometric reflection breaks down, since the
frequency of the incoming radiation nears the plasma frequency of the mirror.
So instead of photons, we will need to use $\sim$1 keV electrons, which do
reflect properly from the mirror surface.

Many photons whose total energy is $E_{T}$ give less position information than
a single photon with energy
\begin{equation}
E_{T}=\frac{hc}{\lambda}.
\end{equation}
So we set
\begin{equation}
E_{T}=\frac{hc}{W}=\frac{hc}{\frac{5}{2}\lambda_{1}},
\end{equation}
and also set $E_{T}$ equal to the thermal output of the mirror during time
$t_{s}=\frac{\lambda_{1}}{\Delta v_{i}}$, the time necessary for the spreading
to reach $\lambda_{1}$:
\begin{equation}
E_{T}=t_{s}e\sigma T^{4}\left[2\pi\left(\frac{5}{4}\lambda_{1}\right)^{2}
\right]
\end{equation}
so
\begin{equation}
\frac{hc}{\frac{5}{2}\lambda_{1}}=\frac{\lambda_{1}}{4\Delta v_{i}}e\sigma
T^{4}\left[2\pi\left(\frac{5}{4}\lambda_{1}\right)^{2}\right],
\end{equation}
which reduces to
\begin{equation}
T^{4} = \frac{1}{t_{s}}\frac{1}{e\sigma}\frac{hc}{\frac{1}{2}\pi\left(\frac{5}
{2}\lambda_{1}\right)^{3}}.
\end{equation}
Since
\begin{equation}
t_{s}=\frac{\lambda_{1}}{\Delta v_{i}}=\frac{4\pi\Delta q_{i}M\lambda_{1}}{h},
\end{equation}
we see that
\begin{displaymath}
T^{4}=\frac{h^{2}c}{e\sigma}\frac{1}{2\pi\Delta q_{i}M\left(\frac{5}{2}\right)
^{3}\lambda_{1}^{4}},
\end{displaymath}

\pagebreak
\noindent
or
\begin{equation}
\label{ind-eq}
T^{4}=\frac{2.6\times 10^{-35}}{\lambda_{1}^{6}\Delta q_{i}}.
\end{equation}
To evaluate (\ref{ind-eq}) we need to make a decision about what value of
$\Delta q_{i}$ to use. This is probably the hardest parameter to pin down
without actually performing the experiment, but the natural choice seems to
be
\begin{equation}
\Delta q_{i}=\frac{1}{2}d,
\end{equation}
which in this case is $1.2\times 10^{-8}$ cm. Then using $\lambda_{1}=10^{-6}$
cm,we get:
\begin{quote}
$\Delta v_{i}=3.8\times 10^{-3}$ cm/s,\\
$t_{s}=2.6\times 10^{-4}$ s,\\
$M=1.1\times 10^{-17}$ g, and\\
$T_{R}=2.2\times 10^2$ K.
\end{quote}

We will subsequently find that there are other temperature limits more
stringent than this. But this limit is conceptually
important, because it is the temperature below which we can treat the mirror as
an independent system.

The main phenomenon that competes with wave packet spreading in determining the
initial velocity is thermal motion {\em within} the mirror. At very low
temperature, thermal energy in the mirror is stored in the lowest frequency
phonons available---the lowest harmonics of the disk.

We can find the temperature cut-off where the two competing effects are roughly
equal by equating the total momenta
\begin{equation}
(\mu v_{s})\sqrt{2}=M\Delta v_{i},
\end{equation}
where $\mu\equiv$ reduced mass $\approx\frac{1}{2}M$. From this, it follows
that
\begin{equation}
\frac{1}{2}\mu v_{s}^{2}=\frac{1}{2}M(\Delta v_{1})^{2}.
\end{equation}
But $\frac{1}{2}\mu v_{s}^{2}$ is half the thermal energy in each mode, which
is also
\begin{equation}
E=\frac{\hbar\omega/2}{e^{\hbar\omega /k_{B}T}-1};
\end{equation}
therefore, at $T_{c}$ we have
\begin{displaymath}
\frac{1}{2}M(\Delta v_{i})^{2}=\frac{\hbar\omega/2}{e^{\hbar\omega /k_{B}T_{c}}
-1}
\end{displaymath}
\begin{equation}
\ln\left[\frac{\hbar\omega}{M(\Delta v_{i})^{2}}+1\right] = \frac{\hbar\omega}
{k_{B}T_{c}},
\end{equation}
so
\begin{equation}
T_{c} = \frac{\hbar\omega}{k_{B}}\left(\ln\left[\frac{\hbar\omega}{M(\Delta
v_{i})^{2}}+1\right]\right)^{-1}.
\label{Tc-anal}
\end{equation}
From the geometry of the problem, we can see that $\omega=\frac{2\pi v_{s}}
{\lambda}$, where $\lambda$ is now the diameter of the mirror disk, so for an
arbitrary value of $\lambda_{1}$,
\begin{equation}
\omega=\frac{2\pi v_{s}}{\frac{5}{2}\lambda_{1}}=\frac{4\pi}{5}\frac{v_{s}}
{\lambda_{1}}=\frac{7.5\times 10^{5}}{\lambda_{1}}.
\end{equation}
So our value of $T_{c}$ becomes
\begin{equation}
T_{c}=(5.7\times 10^{-6})[\lambda_{1}\ln\lambda_{1}(1.1\times 10^{12})]^{-1}.
\end{equation}
For $\lambda_{1}=10^{-6}$ cm,
\begin{equation}
\label{phtemp-eq}
T_{c}=4.1\times 10^{-1} \,{\rm K}.
\end{equation}
Below $T_{c}$, wave packet spreading dominates the effect of thermal phonons
within the mirror.

Now consider the thermal conditions outside the mirror. In order to trap the
mirror in a potential well before the beginning of the experiment proper, the
mirror needs to start with a very small thermal velocity---not too much
greater than $\Delta v_{i}$. Approximating by the ideal gas value,
\begin{equation}
\frac{1}{2}Mv_{T}^{2}=\frac{3}{2}k_{B}T_{g},
\end{equation}
we get
\begin{equation}
v_{T}=6.2\sqrt{T_{g}}\,\, {\rm cm/s}.
\end{equation}
If we set $v_{T}=\alpha\Delta v_{i}$, then
\begin{equation}
\label{gastemp-eq}
T_{g}=3.8\times 10^{-7}\alpha^{2}\, {\rm K}.
\end{equation}
For $\alpha<10^3$, (\ref{gastemp-eq}) is a much tighter restriction than
(\ref{phtemp-eq}). Part of our experimental strategy would be to use a
sequence of trapping maneuvers to raise the allowable values of $\alpha$ and
$T_{g}$. One of these maneuvers would be likely to involve attaching the mirror
to a more massive object using macromolecules that can change their tertiary
structure \cite{ref-terner,ref-alpert}.

We also require that the density of the gas surrounding the mirror be low
enough so that there will be no collisions during the time $t_{s}$. This means
that there must be less than one molecule in the volume
\begin{equation}
V=(v_{g}t_{s})\pi\left(\frac{5}{4}\lambda_{1}\right)^{2},
\end{equation}

\pagebreak
\noindent
where $v_{g}$ is the rms velocity of the gas molecules at temperature $T_{g}$.
For Rubidium (frequently used in Bose-Einstein condensate experiments),
\begin{equation}
v_{g}=1.06\,\alpha \,{\rm cm/s},
\end{equation}
and
\begin{equation}
V=1.33\times 10^{-15}\,\alpha \,{\rm cm}^{3}.
\end{equation}
This yields a density of
\begin{equation}
\rho_{\alpha}=\frac{1.2\times 10^{-6}}{\alpha}\, {\rm mole/L}.
\end{equation}
Note that if $\rho\ll\rho_{\alpha}$, the mirror will sometimes approach the
trap with a Brownian velocity
\begin{equation}
v_{Br}<v_{T}=\alpha\Delta v_{i}.
\end{equation}
Taking full advantage of this, we should be able to work at values of $\alpha>
5$ and hence $T_{g}>10$ $\mu$K.

\section{Conclusions}
\label{concl}
\noindent
We have designed an experiment that must have a consistent outcome. The outcome
can certainly be consistent if desharpening of the $\lambda_{1}$ nodes is not
found to occur. But then quantum mechanics does not correctly predict the
$\lambda_{1}$ pattern. If we want to retain quantum mechanics, $\lambda_{1}$
desharpening should occur. Then, to avoid the Wigner's friend contradiction, we
are forced to jettison SIQM and take another path.

The $\lambda_{1}$ pattern must be unaffected by $\lambda_{2}$ detection, even
though $\lambda_{2}$ detection restricts the mirror to $|z_{m}|\leq\frac
{\lambda_{1}}{4}$. This means that $\lambda_{1}$ develops the desharpened
$|z_{m}|<\lambda_{1}$-related nodes without encountering the mirror in the zone
$\cal{Z}$ where $\lambda_{1}>|z_{m}|>\frac{\lambda_{1}}{4}$. Presumably, this
is pretty much what happens with or without $\lambda_{2}$. $\lambda_{1}$ must
encounter the mirror itself only in a small region $\Delta z\ll\lambda_{1}$,
and in most of the larger region, $|z_{m}|<\lambda_{1}$, it will encounter some
kind of signal from the mirror. This signal cannot reflect the $\lambda_{1}$
wave, but it can modulate the wave to produce the correct interference pattern
if
\begin{quote}
1. the mirror emits the signal continuously, like the wake of a boat, and\\
2. the signal contains enough information about the evolution of the state of
the mirror.
\end{quote}
Note that although the linear wave equation correctly predicts the shape of the
$\lambda_{1}$ nodes, the underlying process, with propagation of the modulating
signal taking the place of wave packet spreading, is fundamentally nonlinear.

Let us now consider the experiment from a mathematical point of view. We find
that there is a ``duality'' of $\lambda_{2}$ position measurement:$\lambda_{1}$
reflection interference. We already know that this duality does not operate in
the usual manner to produce two different sets of basis vectors for the same
space. Instead, the $\lambda_{1}$ and $\lambda_{2}$ measurements lead us to two
different vector spaces, that presumably are tangent in some sense to the
(infinite-dimensional) manifold that actually represents the state of the
system. The superposition principle hold only within these individual vector
spaces.

It is natural to conjecture that this duality is a special case of a
multiplicity of distinct properties and corresponding vector spaces, each of
which is accessed by a different probe of the mirror system.  Ordinarily, each
probe disrupts all the others to such an extent that the multiplicity is not
evident.

Consider the information-carrying signal in this light. The signal exists in
the space that describes the experiment but not in the ``tangent'' spaces that
are the setting for SIQM. The signal is able to correctly modulate
$\lambda_{1}$, because it carries {\em locally} information about phase
correlations that occur elsewhere. Such richness of information content is
possible only if the signal dwells in a very large and profoundly non-linear
space.

The effects we have described can exist only under extreme conditions of low
temperature with very careful state preparation. They are nevertheless based
on rather general features of wave mechanics. The necessary temperature regime
is now accessible to experimentalists. We believe that an experiment of this
type can be performed, although it probably would entail the use of states more
specifically tailored to micro-Kelvin conditions.

\pagebreak
\appendix
\section{Trapping, Confining, and Releasing the Mirror to Start the Experiment}
\noindent
We propose to trap the mirror by floating it in a magnetic field balanced by a
weak fictitious gravitational field due to an acceleration. The magnetic field
would induce a superconduction current in the mirror. The $x$ and $y$
components of the $\vec{B}$ field will then act on the current to produce a
force opposite to the fictitious force. The mirror will sit in a potential well
approximately given by
\begin{equation}
\frac{1}{2}kz_{m}^{2}=\frac{1}{2}Mv^{2}
\end{equation}
In particular, $k$ must satisfy
\begin{equation}
\frac{1}{2}k(\Delta q_{i})^{2}=\frac{1}{2}M(\Delta v_{i})^{2},
\end{equation}
so
\begin{equation}
k=1.1\times 10^{-6} \,{\rm erg/cm}^{2}
\end{equation}
and the oscillator frequency $\omega_{os}$ is
\begin{equation}
\omega_{os}=\sqrt{\frac{k}{M}}=3.2\times 10^{5}.
\end{equation}

Let us calculate the magnitude of the $\vec{B}$ field. We simplify the problem
by replacing the mirror disk with a ring of radius $\lambda_{1}$. Also,
\begin{equation}
B_{z}\approx B_{0}\sin\omega t,
\end{equation}
where $\frac{2\pi}{\omega}\gg t_{s}$, and
\begin{displaymath}
B_{r}\approx B_{\perp}\sin\omega t
\end{displaymath}
\begin{equation}
B_{\perp}\approx \eta B_{0},
\end{equation}
where $\eta$ is slowly varying with respect to $z$. Then from
\begin{equation}
\oint E\cdot ds = -\frac{1}{c}\frac{d\Phi_{B}}{dt}
\end{equation}
we get
\begin{equation}
E(2\pi\lambda_{1})=-\frac{1}{c}B_{0}\pi\lambda_{1}^{2}(\omega\cos\omega t)
\end{equation}
\begin{equation}
E=-\frac{1}{2c}\lambda_{1} B_{0}\omega\cos\omega t.
\end{equation}
This acts on superconducting electrons to produce
\begin{equation}
a=\frac{F}{m_{e}}=\frac{eE}{m_{e}}=-\frac{e\lambda_{1}}{2m_{e}c}B_{0}\omega
\cos\omega t,
\end{equation}
and
\begin{equation}
v=-\frac{e\lambda_{1}}{2m_{e}c}B_{0}\sin\omega t.
\end{equation}
The the radial component of $\vec{B}$, $B_{r}$ acts on each electron to produce
a force
\begin{displaymath}
F_{e}=-\frac{ev}{c}B_{r}
\end{displaymath}
\begin{equation}
F_{e}=\frac{e^{2}\lambda_{1}}{2m_{e}c^{2}}B_{0}B_{\perp}\sin^{2}\omega t.
\end{equation}
If there are $\nu$ Cooper pairs for each lattice position, and $N$ is the total
number of atoms in the mirror, then the total upward force on the mirror is
\begin{equation}
F_{EM}=N\nu\eta\frac{e^{2}\lambda_{1}}{m_{e}c^{2}}B_{0}^{2}\sin^{2}\omega t.
\end{equation}

Because $t_{EM}=\frac{2\pi}{\omega}\gg t_{s}$, we can time the experiment to
be performed when $F_{EM}$ is at its maximum,
\begin{equation}
F_{EM-max}=N\nu\eta\frac{e^{2}\lambda_{1}}{m_{e}c^{2}}B_{0}^{2}.
\end{equation}
The balancing acceleration would have the same periodicity as $F_{EM}$.
The potential well is created by the inhomogeneity inhomogeneity of the
$\vec{B}$ field as a function of $z$.
\begin{equation}
H_{total}=\int F_{EM} dz - F_{accel}z.
\end{equation}
So we have
\begin{displaymath}
H_{total-max}=\int F_{EM-max} dz - F_{accel-max}z
\end{displaymath}
\begin{equation}
H_{total}\approx\frac{1}{2}\frac{\partial F_{EM-max}}{\partial z}z^{2},
\end{equation}
so we can see that
\begin{equation}
k\approx\frac{\partial F_{EM}}{\partial z}\approx 2N\nu\eta\frac{e^{2}\lambda
_{1}}{m_{e}c^{2}}B_{0}\frac{\partial B_{0}}{\partial z}.
\end{equation}

\pagebreak
\noindent
Therefore, $B_{0}\frac{\partial B_{0}}{\partial z}\approx\frac{km_{e}c^{2}}
{2N\nu\eta e^{2}\lambda_{1}}\approx\frac{1.8\times 10^{7}}{\nu\eta}$. If we
set $\nu=10^{-3}$ and $\eta=10^{-1}$, then $B_{0}\frac{\partial B_{0}}{\partial
z}\approx1.8\times 10^{11}$. If $B_{0}$ varies by about 25 percent over the
distance $\lambda_{1}$ in the vicinity of the mirror, then
\begin{displaymath}
\frac{\partial B_{0}}{\partial z}\lambda_{1}|_{z=0}\approx\frac{1}{4}B_{0}|
_{z=0}
\end{displaymath}
\begin{equation}
B_{0}\frac{\partial B_{0}}{\partial z}|_{z=0}\approx\frac{B_{0}^{2}}{4
\lambda_{1}}.
\end{equation}
This reduces to $B_{0}|_{z=0}\approx 8.5\times 10^{2}$ G.

Finally, the mirror is released by turning off the field over a time scale
$t_{R}$ that satisfies
\begin{equation}
t_{R}\ll t_{s}.
\end{equation}
Ideally, one would also want
\begin{equation}
t_{R}<\frac{2\pi}{\omega_{os}}.
\end{equation}
The mirror would be briefly exposed to a strong $\vec{E}$ field during the
turn-off.

\pagebreak

\begin{figure}
\caption{The Bedford-Wang experiment. D detects the trigger photon
($\psi_{TP}$), then sends a signal to the mechanical linkage (ML), which moves
the slit covers from $\ket{Ab}$ to $\ket{aB}$.}
\label{BWexpt}
\end{figure}

\begin{figure}
\caption{Apparatus shown in A open, B closed configuration (state $\ket{Ab}$).
Note that the $\lambda_{2}$ beam and detector can be tilted slightly out of
the $xy$-plane to avoid confusion.}
\label{prelim}
\end{figure}

\begin{figure}
\caption{The apparatus modified for a long excursion taken by the $\lambda_{2}$
photons.}
\label{pathdelay}
\end{figure}

\begin{figure}
\caption{Time-line of the Gedanken experiment.}
\label{timeline}
\end{figure}

\begin{figure}
\caption{A bifurcated mirror generating a simple interference pattern. The
$z_{1}$ reflected wave and the $z_{2}$ reflected wave are out of phase by $\pi$
at the center of the pattern.}
\label{bifur}
\end{figure}

\begin{figure}
\caption{Reflection off a mirror of diameter $\frac{5}{2}\lambda_{1}$. The two
paths arrive at the interference screen out of phase by $\pi$; however, they do
not exactly cancel, because the the amplitude decays along paths of differing
lengths. To avoid confusion between the outgoing and reflected $\lambda_{1}$
quanta, the beam is tilted slightly out of the $xz$-plane.}
\label{pl-fig1}
\end{figure}

\begin{figure}
\caption{The range of paths that the reflected quanta may take, for a mirror
with a spread of $\Delta z=\lambda_{1}$.}
\label{pl-fig2}
\end{figure}

\begin{figure}
\caption{Interference patterns for a mirror with a Gaussian wave packet with
$\Delta z=\lambda_{1}$ and one for which the same Gaussian has been truncated
to $\pm\frac{\lambda_{1}}{4}$.}
\label{graphs1}
\end{figure}

\begin{figure}
\caption{The configuration for the non-null experiment. Shown are the
$\lambda_{1}$ and $\lambda_{2}$ paths corresponding to the extreme mirror
positions permitted by a D$_{\lambda_{2}}$ detection event.}
\label{real}
\end{figure}


\begin{references}

\bibitem{ref-bw1}D.~Bedford, D.~Wang, Found. of Phys. {\bf 6} 599 (1976).
\bibitem{ref-bw2}D.~Bedford, D.~Wang, Nuovo Cimento B {\bf 32} 243 (1976).
\bibitem{ref-bw3}D.~Bedford, D.~Wang, Nuovo Cimento B {\bf 26} 313 (1975).
\bibitem{ref-bw4}D.~Bedford, D.~Wang, Nuovo Cimento B {\bf 37} 55 (1977).
\bibitem{ref-bw5}D.~Bedford, D.~Wang, Found. of Phys. {\bf 13} 987 (1983).
\bibitem{ref-guthk}D.~Guthkovski, M.~V.~Valdes Franco, Found. of Phys. {\bf 13}
963 (1983).
\bibitem{ref-lwang1}L.~J.~Wang, X.~Y.~Zou, L.~Mandel, J. Opt. Soc. Am. B
{\bf 8} 978 (1991).
\bibitem{ref-lwang2}L.~J.~Wang, X.~Y.~Zou, L.~Mandel, Phys. Rev. A {\bf 44}
4614 (1991).
\bibitem{ref-lwang3}L.~J.~Wang, X.~Y.~Zou, L.~Mandel, Phys. Rev. Lett. {\bf 67}
318 (1991).
\bibitem{ref-ou1}Z.~Y.~Ou, L.~J.~Wang, L.~Mandel, Phys. Rev. A {\bf 40} 1428
(1989).
\bibitem{ref-ou2}Z.~Y.~Ou, L.~J.~Wang, L.~Mandel, Phys. Rev. A {\bf 41} 1597
(1990).
\bibitem{ref-terner}J.~Terner, J,~D.~Stong, T.~G.~Spiro, M.~Nagumo,
M.~F.~Nicol, M.~A.~El-Sayed, in {\em Hemoglobin and Oxygen Binding}, edited by
C. Ho (Elsevier, New York, 1982).
\bibitem{ref-alpert}B.~Alpert, L.~Lindqvist, S.~El~Mohsni, F.~Tfibel, in
{\em Hemoglobin and Oxygen Binding}, edited by C. Ho (Elsevier, New York,
1982).

\end{references}
\end{document}